\documentclass{svmult}
\usepackage{makeidx}         % allows index generation
\usepackage{graphicx}        % standard LaTeX graphics tool
\usepackage{multicol}        % used for the two-column index
\usepackage[bottom]{footmisc}% places footnotes at page bottom
\makeindex             % used for the subject index
\begin{document}

\title*{Recurrence analysis of the NASDAQ crash of
April 2000 }
\author{Annalisa Fabretti\inst{1}\and
Marcel Ausloos\inst{2}}
\institute{Department of Mathematics for Economy, Insurances and
Finance Applications, University of Roma 1, La Sapienza I-00100
Rome, Italy \texttt{annalisa.fabretti@uniroma1.it} \and SUPRATECS,
B5, University of Li$\grave e$ge, B-4000 Li$\grave e$ge, Euroland
\texttt{marcel.ausloos@ulg.ac.be}}
\maketitle

\begin{abstract} Recurrence Plot (RP) and Recurrence Quantification Analysis
(RQA) are signal
    numerical analysis methodologies able to work with non linear dynamical
systems and non stationarity. Moreover they well evidence changes
in the states of a dynamical system.  It is shown that RP and RQA
detect the critical regime in financial indices (in analogy with
phase transition) before a bubble bursts, whence allowing to
estimate the bubble initial time. The analysis is made on NASDAQ
daily closing price between Jan. 1998 and Nov. 2003. The NASDAQ
bubble initial time has been estimated to be  on Oct. 19, 1999.
\end{abstract} %

\section{Introduction}
\label{sec:1}
 Recent papers have shown some analogy between crashes
and phase transitions \cite{sornette, feigenbaum, aus3}; like in
earthquakes, log periodic oscillations have been found before some
crashes \cite{aus1, joh_and_sorn}, then it was proposed that an
economic index $y(t)$ increases as a complex power law, whose
first order Fourier representation is
\begin{equation}\label{log}
   y(t)=A+B\ln(t_{c}-t)\{1+C\cos[\omega\ln(t_{c}-t)+\phi]\}
\end{equation}
where $A$, $B$, $C$ , $\omega$, $\phi$ are constants and $t_{c}$
is the critical time (rupture time).

An endogenous crash is preceded by an unstable phase where any
information is amplified; this critical period takes the name of
'speculative bubble'.

Recurrence Plots are graphical tools elaborated by Eckmann,
Kamphorst and Ruelle in 1987 and are based on Phase Space
Reconstruction \cite{rp}. In 1992, Zbilut and Webber \cite{crq}
proposed a statistical quantification of RPs and gave it the name
of 'Recurrence Quantification Analysis' (RQA).
 RP and RQA are good in working with non stationarity and noisy data,
  in detecting changes in data behavior, in particular in
detecting breaks, like a phase transition \cite{lambertz},  and in
informing about other dynamic properties of a time series
\cite{rp}. Most of the applications of RP and RQA are at this time
in the field of physiology and biology, but some authors have
already applied these
  techniques to financial data \cite{anto_and_vorlow,holyst2}.
 We have used RP and RQA techniques
for detecting critical regimes preceding an endogenous crash seen
as a phase transition and whence give an estimation of the initial
bubble time.

It has been simulated a signal as in \ref{log}, the analysis is made
on  NASDAQ, taken over a time span of
     6 years including the known crash of April 2000
\cite{joh_and_sorn}. The
    series are also divided into subseries in order to investigate
    changes in the evolution of the signal. Then the RPs of all time
    series have been observed, compared and discussed. This work
    is extracted by \cite{AF_and_MA}.

\section{Recurrence Analysis} \label{sec:2}
 The changing state of
a dynamic system can be indeed represented by sequences of 'state
vectors' in the phase space. Each unknown point of the phase space
at time $i$ is reconstructed by the delayed vector
$y(i)={X_i,X_{i+d},...,X_{i+(m-1)d}}$ in an $m$-dimensional space.
\paragraph{Recurrence Plot} The Recurrence Plot (RP) is a matrix
of points $(i,j)$ where each point is said to be recurrent and
marked with a dot if the distance between the delayed vectors
$y(i)$ and $y(j)$ is less than a given threshold $\epsilon$. As
each coordinate  $i$ represents a point in time, RP provides
information about the temporal correlation of phase space points
\cite{rp}.

Therefore RPs can be used to test a system deterministic behavior
through the percentage of recurrent points belonging to parallel
lines. In fact for a periodic or a deterministic signal patterns
like parallel lines appear.
  In so doing, RPs are useful tools
for the preprocessing of experimental time series and provide a
comprehensive image of the dynamic course at a glance
\cite{lambertz}.
\paragraph{Recurrence Quantification Analysis} RQA quantifies the presence of patterns, like
parallel lines of RPs, with 5 RQA variables: the percentage of
recurrent points (\%REC). The percentage of recurrent points
forming line segments
   parallel to the main diagonal (\%DET). The longest line segment measured parallel to
   the main diagonal (MAXLINE). The slope of line-of-best-fit through \%REC as a
function of the displacement from the main
   diagonal (excluding the last 10\% range) (TREND). The Shannon entropy of the distribution
   of the length of line segments parallel to the main diagonal (ENT).

  \section{Analysis and Conclusions} 
\begin{figure}
\centering
  \includegraphics[scale=0.5]{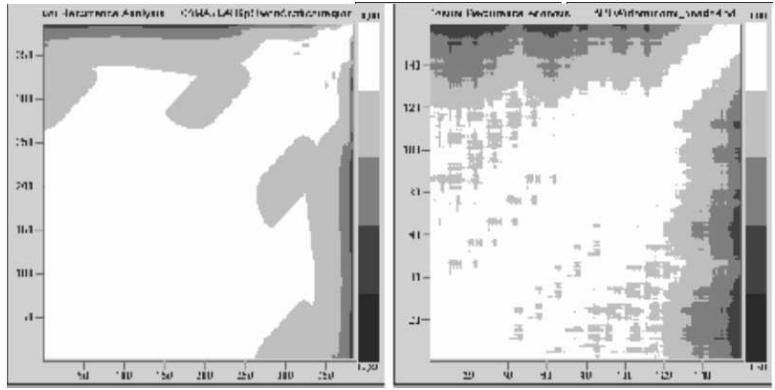}
  \caption{(lhs) RP of a log periodic signal as generated by equation (\ref{log}).
  The arrow shape on the lhs plot is the sign of a strong trend;
the curve lines are due to the log periodicity. (rhs) RP of the
NASDAQ subseries (IV) from Jan., 1999 to Dec., 1999. It is worth
to note the similarity between these two RPs}\label{FigLog}
\end{figure}

\begin{figure}
\centering
  \includegraphics[height=4cm]{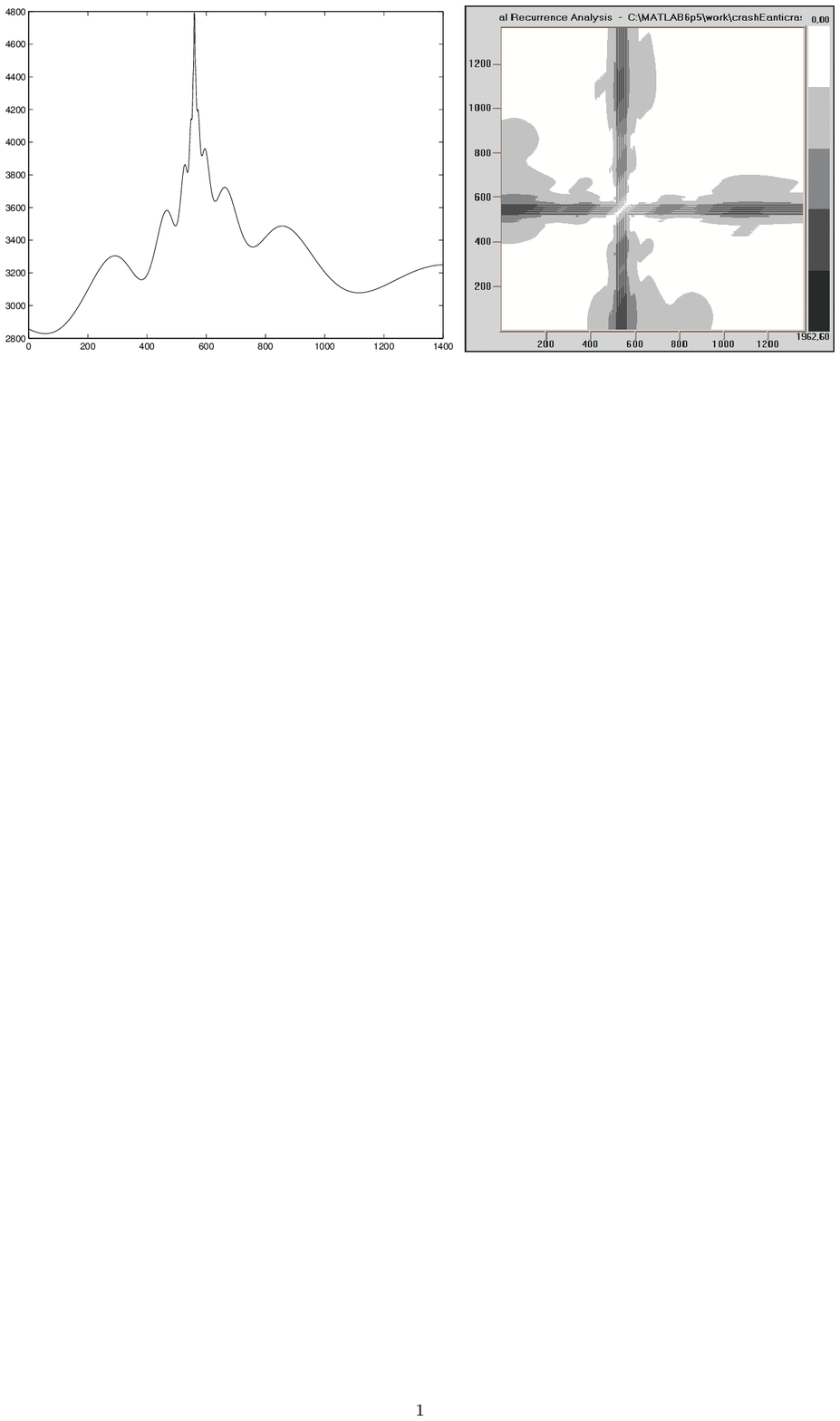}
  \caption{(RHS) the RP of a simulated phase transition of a signal (LHS) following the law (\ref{log}) before and after the critical event.}\label{FigCrash}
\end{figure}

  In order
to study the crash from the point of view of a phase transition
with log periodic precursors, a log periodic signal, generated by
equation (\ref{log}), has been simulated, its RP is shown in Fig.
\ref{FigLog}(lhs). The 'arrow' shape is due to the trend, the not
smooth border ('color') lines are due to the log periodicity. It
has been also considered a phase transition signature. In Fig.
\ref{FigCrash}(lhs) an arbitrary signal is plotted before and
after a peak, taking into account the anti-bubble phenomenon after
a crash \cite{antibubble}. The RP aspect of Fig.
\ref{FigCrash}(RHS) reveals a feature far from the normal signal
evolution; to be noted the well marked black bands corresponding
to the crash time.

  About NASDAQ after studied the whole time series the data
   has been divided into subseries of 200 days, overlapping each other of almost 5 months,
   in order to further analyze whether and how the data changes.

\begin{figure}
\centering
  \includegraphics[height=4cm]{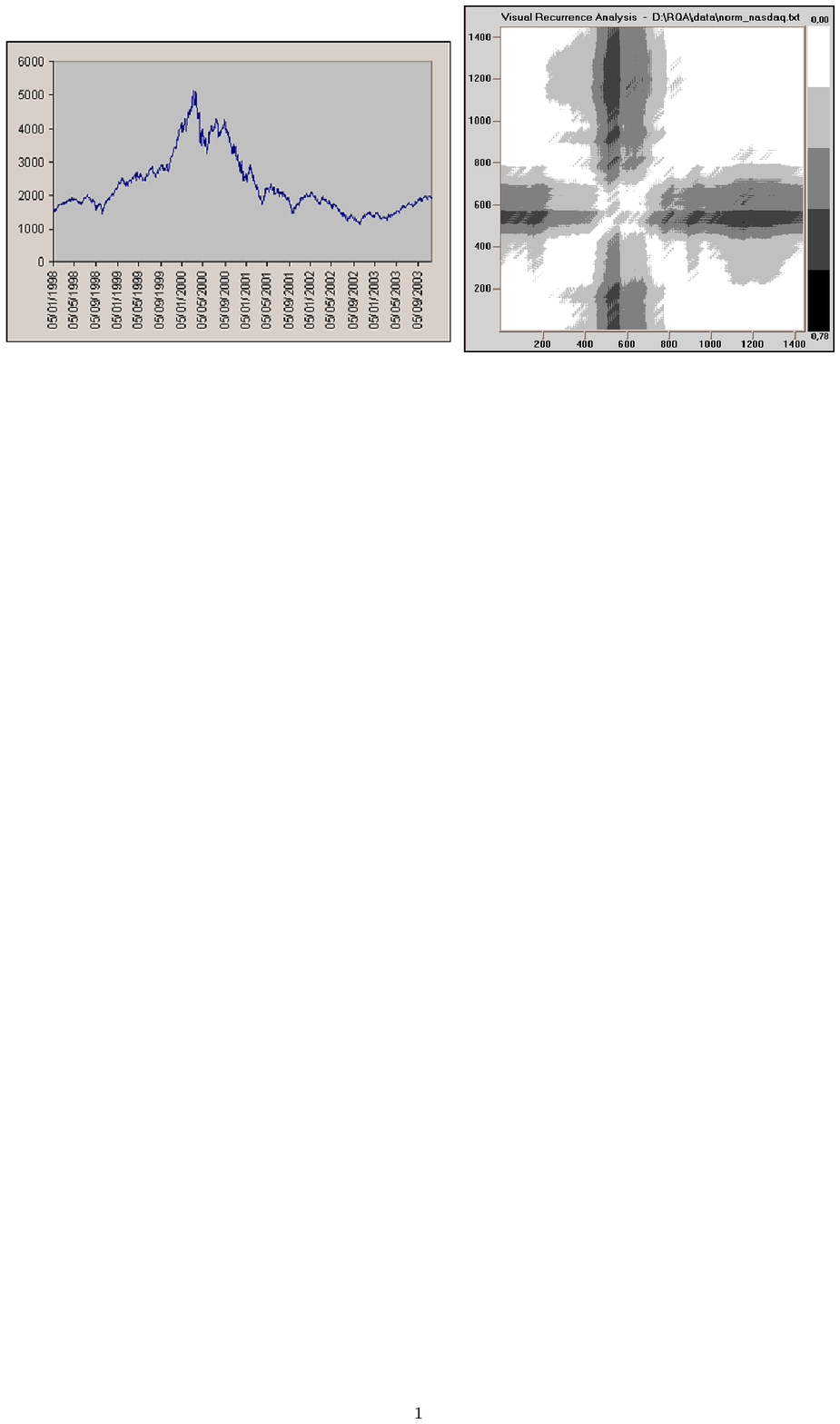}
  \caption{(LHS)daily closing price of NASDAQ from Jan. 05, 1998 to Nov. 21, 2003; (RHS) RP of  NASDAQ from Jan. 05, 1998 to Nov. 21, 2003.
   The dark grey band delimited by horizontal coordinates $x=504$ and
$x=566$, encircled by a grey area delimited by horizontal
coordinates $x=452$ and $x=690$, is the 'image' of the crash of
April 2000.
  It is a 'strong event' but afterwards the normal regime is restored}\label{FigG}
\end{figure}

Fig. \ref{FigG}(rhs) is the RP of NASDAQ between Jan. 05, 1998 and
Nov. 21, 2003. Of interest is the dark grey vertical band
surrounded by a lighter grey area, delimited by horizontal
coordinates $x=452$ and $x=690$ corresponding to Oct. 19, 1999 and
Sept. 27, 2000. In correspondence of the period in which the
bubble grows, RQA variables take the highest absolute values
\cite{AF_and_MA}.

 It is worth to note the same RP shape of the
phase transition signature in Fig. \ref{FigCrash}. Considering
that each coordinate in RP is linked with the time series, the
border line of a grey or black band reveals the time when the data
behavior starts to change. Noting that the dates here above fall
in the same time interval as the bubble and the subsequent crash,
it can be supposed that the initial bubble time occurs at $x= 452$
(Oct. 19,1999). We can thus deduce that on such a day the
evolution of the system changes, i.e. the evolution passes from a
normal regime to a critical regime.  This is an a posteriori
estimation of the initial bubble time, but through the analysis of
the subseries one can argue to be able to recognize the beginning
of the bubble with some delay before the bubble grows. In fact,
while the RPs of the first (I), the second (II) and of the third
(III) subseries do not present any remarkable pattern \cite{AF_and_MA}
(they are quite homogeneous except some local maximum reached by
the index) the fourth subseries presents an interesting pattern:
the RP in Figure \ref{FigLog}(rhs) shows the characteristic shape
typical of the strong trend of a speculative bubble as studied and
pointed out in Fig.
 \ref{FigLog}(lhs). The trend starts to be
significant in the middle of Oct. 1999. This indicates that the RP
has changed indeed when the bubble has started.  Even the RQA
variables, in Table \ref{tableRqaNasdaq}, evidence in this period
the highest values. It has to be underlined that this IV period
does not include the crash time, but stops in Dec. 1999 before the
bubble bursts. Even in the fifth (V) subseries RP,
 the bubble
beginning is not so evidently as it was in the fourth subseries.

In conclusion it has been shown that, with some delay  as respect
to the beginning but enough time before the crash (3 months in
this particular case), such that a warning could be given, RP and
RQA detect a difference in state and recognize the critical
regime.

\index{paragraph}
%\clearpage

\begin{table}
\centering \caption{RQA of NASDAQ on 5 the subseries studied of
200 days.}\label{tableRqaNasdaq}
\begin{tabular}{llllll}
\hline\noalign{\smallskip}
   Subseries & I & II & III & IV& V\\
   periods &Jan.1998 & June
   1998&Oct.1998&Feb.1999&July
   1999 \\
          &Oct.1998&Feb.1999&July1999&Dec.1999&May2000\\
   \noalign{\smallskip}\hline\noalign{\smallskip}
\%REC & 6.075 & 9.141 & 5.513 & 17.146 & 9.246 \\
  \%DET & 35.980 & 36.119 & 29.079 & 45.018 & 54.511 \\
  MAXLINE & 125 & 158 & 83 & 179 &166\\
  ENT& 2.522 & 3.547 & 2.585  & 4.054& 3.301 \\
  TREND(units/1000points) & -105.125 & -155.824 & -97.808 & -273.775 & -138.501 \\
  \noalign{\smallskip}\hline
\end{tabular}
%%\end{threeparttable}
\end{table}

\printindex

\end{document}